\documentclass[12pt,a4paper]{article}

\usepackage[utf8]{inputenc}
\usepackage[russian]{babel}
\usepackage{fullpage}

\usepackage{amsmath}
\usepackage{amssymb}
\usepackage{amsthm}
\usepackage{fixmath}

\title{Краткое изложение $\lambda$-исчисления}
\author{Антон Салихметов}

\begin{document}

\maketitle

\begin{abstract}
Данный текст представляет собой чрезвычайно сжатый конспект классической монографии по $\lambda$-исчислению~\cite{lambda}.
Он может быть интересен тем, кто планировал взяться за систематическое изучение данной темы, уже в общих чертах ознакомившись с ней, но откладывал из-за сложной структуры основной монографии, определения и основные результаты в которой довольно разрозненны.
Здесь мы попытаемся сделать изложение, напротив, абсолютно линейным, и, конечно, несравнимо более коротким, избегая лишних определений и примеров, а сосредоточившись на необходимых терминологии, обозначениях и утверждениях, которые, в свою очередь, изложены близко к оригинальному тексту.

Мы начнем с определения системы $\lambda\beta\eta$, то есть классического бестипового экстенсионального $\lambda$-исчисления.
Затем перейдем к комбинаторной логике, теореме о неподвижной точке и синтаксическому сахару.
Наконец, заключительная часть конспекта~---~построение топологии на выражениях этой системы, призванной объяснить кажущееся противоречие: отображения множества выражений в себя содержатся в самом этом множестве при его счетности.
На самом же деле, множество наделяется надлежащей топологией, в которой выражения представляют собой непрерывные отображения.
\end{abstract}

\section{Теория}

Множество $\lambda$-\textit{выражений} $\Lambda$ строится индуктивно из \textit{переменных}
$$
x, y, z \ldots \in \Lambda
$$
с помощью \textit{абстракций}
$$
M \in \Lambda \Rightarrow \lambda x.M \in \Lambda
$$
и \textit{аппликаций}
$$
M, N \in \Lambda \Rightarrow M\ N \in \Lambda,
$$
при этом аппликация лево-ассоциативна:
$$
(M) \equiv M, \quad M\ N\ P \equiv (M\ N)\ P.
$$

Рефлексивное транзитивное отношение $M \subset N$ означает, что $M$ является \textit{подвыражением} выражения $N$:
\begin{align*}
M \subset M &\subset \lambda x.M; \\
M &\subset M\ N \supset N;\\
M \subset N \land N \subset P \Rightarrow M &\subset P.
\end{align*}

$\text{FV}(M)$ — это множество \textit{свободных} переменных в выражении $M$:
\begin{align*}
\text{FV}(x) &\equiv \{x\}; \\
\text{FV}(\lambda x.M) &\equiv \text{FV}(M) \setminus \{x\}; \\
\text{FV}(M\ N) &\equiv \text{FV}(M) \cup \text{FV}(N).
\end{align*}

Переменные, которые не являются свободными, называются \textit{связанными} и могут быть заменены другой переменной (такое преобразование называют $\alpha$-конверсией):
$$
y \not\in \text{FV}(M) \Rightarrow \lambda x.M \equiv \lambda y.M[x := y], 
$$
где $M[x := N]$~---~результат \textit{подстановки}, определяемый следующим образом.

\begin{enumerate}
\item $x[x := P] \equiv P$.
\item $y[x := P] \equiv y$.
\item $(\lambda y.M)[x := P] \equiv \lambda y.M[x := P]$.
\item $(M\ N)[x := P] \equiv M[x := P]\ N[x := P]$.
\end{enumerate}

В третьем пункте не нужно специально оговаривать условие
$$
x \not\equiv y \land y \not\in \text{FV}(P),
$$ так как оно выполняется в силу соглашения о переменных: если в определенном математическом контексте встречаются термы $M_1, \dots , M_n$, то подразумевается, что связанные переменные в них выбраны так, чтобы они были отличны от свободных переменных.

Если множество $\text{FV}(M)$ пусто, то $M$ называют \textit{комбинатором}. Множество всех комбинаторов обозначают $\Lambda^0$:
$$
\Lambda^0 \equiv \{M \in \Lambda\ |\ \text{FV}(M) = \varnothing\}.
$$

Следующие отношения $\beta$, $\eta$ и $\beta\eta$ являются \textit{редукциями}:
\begin{align*}
\beta &\equiv \{((\lambda x.M)\ N, M[x := N])\ |\ M, N \in \Lambda\}; \\
\eta &\equiv \{(\lambda x.M\ x, M)\ |\ M \in \Lambda, x \not\in \text{FV}(M)\}; \\
\beta\eta &\equiv \beta \cup \eta.
\end{align*}

Выражение, подвыражением которого является дырка $(\phantom M)$, называется \textit{контекстом} и обозначается $C[\phantom M]$, при этом $C[M]$ — результат подстановки выражения $M$ вместо дырки в контексте $C[\phantom M]$.

Если $\sigma$~---~редукция, то выражение $M$~---~$\sigma$-\textit{редекс}, если $\exists N: (M, N) \in \sigma$.
Также можно говорить и о $\sigma$-\textit{конверсии} <<$=_\sigma$>>:
\begin{align*}
(M, N) \in \sigma \Rightarrow C[M] &\rightarrow_\sigma C[N]; \\
M &\twoheadrightarrow_\sigma M; \\
M \rightarrow_\sigma N \Rightarrow M &\twoheadrightarrow_\sigma N; \\
M \twoheadrightarrow_\sigma N \land N \twoheadrightarrow_\sigma P \Rightarrow M &\twoheadrightarrow_\sigma P; \\
\exists P: M \twoheadrightarrow_\sigma P \land N \twoheadrightarrow_\sigma P \Rightarrow M &=_\sigma N.
\end{align*}

$\sigma$-\textit{нормальной формой} называют выражение $M$, если $\nexists N: M \rightarrow_\sigma N$.
В экстенсиональном $\lambda$-исчислении под \textit{редексом} имеют в виду $\beta\eta$-редекс, а под \textit{нормальной формой} — $\beta\eta$-нормальную форму.
Говорят, что $M$ \textit{имеет} нормальную форму $N$, если $M \twoheadrightarrow N$.
При этом $\beta\eta$-конверсию обычно обозначают просто «$=$», и это неслучайно: формально система $\lambda\beta\eta$ является эквациональной теорией.
Так как такие теории свободны от логики, непротиворечивость в них определяется несколько иначе.

Равенством будем считать формулу вида $M = N$, где $M$, $N$~---~$\lambda$-выражения; такое равенство \textit{замкнуто}, если $M$ и $N$~---~комбинаторы.
Пусть $T$~---~формальная теория, формулами которой являются равенства.
Тогда говорят, что $T$ \textit{непротиворечива} (и пишут $\text{Con}(T))$, если в $T$ доказуемо не любое замкнутое равенство.
В противном случае говорят, что $T$ \textit{противоречива}.

Одна из причин рассмотрения $\lambda\beta\eta$ состоит в том, что эта теория обладает определенным свойством полноты: для всех комбинаторов $M$ и $N$, имеющих нормальную форму, либо $M = N$, либо $\neg\text{Con}(\lambda\beta\eta + (M = N))$.

\textit{Стратегия}~---~это такое отображение $F: \Lambda \rightarrow \Lambda$, что $\forall M: M \twoheadrightarrow F(M)$.
Для \textit{одношаговой} стратегии выполняется $M \rightarrow F(M)$, если $M$ не является нормальной формой.
Стратегия называется \textit{нормализующей}, если для любого выражения $M$, имеющего нормальную форму $N$, для некоторого числа $n$ выполняется $F^n(M) \equiv N$.
\textit{Левая редукция} $F_l$~---~одна из самых простых одношаговых нормализующих стратегий: она заключается в выборе $\beta$-редекса, значек «$\lambda$» в котором стоит текстуально левее, чем у других $\beta$-редексов, либо левого $\eta$-редекса, если $\beta$-редексов нет.

Таким образом, если два терма имеют общую нормальную форму, то с помощью левой редукции доказательство соответствующего равенства можно получить за конечное число простых шагов.
Если же формула недоказуема, то либо процесс не завершается вовсе, либо он завершается на разных нормальных формах.

\section{Сахар}

Множество комбинаторов $\Xi$ \textit{порождает} наименьшее множество $\Xi^+$ как замыкание по аппликации:
\begin{align*}
\Xi &\subseteq \Xi^+; \\
M, N \in \Xi^+ \Rightarrow M\ N &\in \Xi^+.
\end{align*}

Множество $\Xi$ называется \textit{базисом}, если $\forall M \in \Lambda^0: \exists N \in \Xi^+: M = N$.

Произвольную абстракцию можно смоделировать с помощью $S$ и $K$:
\begin{align*}
S \equiv \lambda x.\lambda y.\lambda z.x\ z\ (y\ z)&; \\
K \equiv \lambda x.\lambda y.x&; \\
I \equiv \lambda x.x& = S\ K\ K; \\
x \not\in \text{FV}(P) \Rightarrow \lambda x.P& = K\ P; \\
\lambda x.P\ Q& = S\ (\lambda x.P)\ (\lambda x.Q).
\end{align*}

Следовательно, комбинаторы $K$ и $S$ задают базис.
Произвольный комбинатор $M$ зачастую описывают не в виде $\lambda$-выражения, а с помощью аксиом. Например, формальная система \textit{комбинаторной логики} $\text{CL}$ определяется двумя аксиомами:
\begin{align*}
K\ P\ Q &= P; \\
S\ P\ Q\ R &= P\ R\ (Q\ R).
\end{align*}

Существуют и одноточечные базисы: один из таких базисов задает комбинатор
$$
X \equiv \lambda x.x\ K\ S\ K.
$$
Действительно, легко проверить, что $X\ X\ X = K$ и $X\ (X\ X) = S$.

Стандартными комбинаторами считаются не только составляющие некоторый базис для комбинаторной логики, но и многие другие полезные $\lambda$-выражения.
Одним из первых примеров обычно дают простейший комбинатор, не имеющий нормальной формы:
$$
\Omega \equiv \omega\ \omega, \quad \omega \equiv \lambda x.x\ x.
$$

Далее, истинностные значения $T \equiv K$ и $F \equiv \lambda x.I$ позволяют обозначить выражением $B\ M\ N$ операцию <<если $B$, то $M$, иначе $N$>>.
Действительно: если $B = T$, то выражение равно $M$; если $B = F$, то выражение равно $N$.
Если $B$ отличен от $T$ и $F$, то результат может быть произвольным.

Как и в теории множеств, в $\lambda\beta\eta$ можно определить упорядоченные пары:
$$
[M, N] \equiv \lambda x.x\ M\ N, \quad [M, N]\ T \twoheadrightarrow M, \quad [M, N]\ F \twoheadrightarrow N.
$$

\textit{Цифровая система}~---~это последовательность комбинаторов $\lceil 0 \rceil, \lceil 1 \rceil, \lceil 2 \rceil \dots$, для которой существуют \textit{следование} $S^+$ и \textit{проверка на нуль} $\text{Zero}$:
\begin{align*}
S^+\ \lceil n \rceil &= \lceil n + 1 \rceil; \\
\text{Zero} \lceil 0 \rceil &= T; \\
\text{Zero} \lceil n + 1 \rceil &= F.
\end{align*}

В \textit{стандартной} цифровой системе выбраны
\begin{align*}
\lceil 0 \rceil &\equiv I; \\
S^+ &\equiv \lambda x.[F, x]; \\
\text{Zero} &\equiv \lambda x.x\ T.
\end{align*}

Цифровая система называется \textit{адекватной}, если относительно нее определимы все рекурсивные функции.
Для выполнения этого свойства достаточно, чтобы нашлась \textit{функция предшествования} $P^-$.
Для стандартной цифровой системы это комбинатор
$$
P^- \equiv \lambda x.x\ F.
$$

Одним из основных результатов $\lambda$-исчисления является теорема о неподвижной точке: для любого $F$ существует $X$, такой, что $F\ X = X$.
Ее доказательство конструктивно.
Пусть $W \equiv \lambda x.F\ (x\ x)$ и $X \equiv W\ W$.
Тогда имеем
$$
X \equiv (\lambda x.F\ (x\ x))\ W = F\ (W\ W) = F\ X,
$$
что и требовалось доказать.
Читатель, возможно, заметил одну особенность в доказательстве этой теоремы.
Чтобы установить, что $F\ X = X$, мы начинаем с терма $X$ и редуцируем его к $F\ X$, а не наоборот.

\textit{Комбинатор неподвижной точки}~---~это терм $Y$, такой, что для любого $F$ имеет место $Y\ F = F\ (Y\ F)$, то есть $Y\ F$~---~неподвижная точка для $F$.
Заметим, что свойство $Y\ F \twoheadrightarrow F\ (Y\ F)$ не имеет места в общем случае.
Поэтому полезен обладающий таким свойством комбинатор неподвижной точки, принадлежащий Тьюрингу:
$$
\Theta \equiv (\lambda x.\lambda y.y\ (x\ x\ y))\ (\lambda x.\lambda y.y\ (x\ x\ y)).
$$

Комбинатор неподвижной точки позволяет решать задачи следующего типа: построить $F$, такой, что
$$
F\ x\ y = F\ y\ x\ F.
$$
Действительно, решение оказывается несложным:
$$
F\ x\ y = F\ y\ x\ F
$$
следует из равенства
$$
F = \lambda x.\lambda y.F\ y\ x\ F,
$$
а оно вытекает из
$$
F = (\lambda f.\lambda x.\lambda y.f\ y\ x\ f)\ F.
$$
Теперь положим
$$
F \equiv Y\ (\lambda f.\lambda x.\lambda y.f\ y\ x\ f)
$$
или, еще лучше,
$$
F \equiv \Theta\ (\lambda f.\lambda x.\lambda y.f\ y\ x\ f),
$$
и все в порядке.

\section{Топология}

Введем некоторые обозначения.

Во-первых, \textit{металамбда-абстракция} $\mathbold\lambda x.f(x)$~---~безымянная запись теоретико-множественной функции $f$, например $(\mathbold\lambda x.x^2 + 1)(3) = 10$.

Во-вторых, определим множество \textit{кодов} конечных последовательностей (в какой-либо стандартной их кодировке натуральными числами)
$$
\text{Seq} = \{\langle n_1, \dots , n_k \rangle\ |\ k \in \mathbb N, n_1, \dots , n_k \in \mathbb N\} \cup \{\langle\phantom M \rangle\}
$$
и следующие полезные обозначения для них.
\begin{itemize}
\item \textit{Длина}~---~количество элементов в последовательности:
\begin{align*}
\text{lh}(\langle \phantom M \rangle) &= 0; \\
\alpha = \langle n_1, \dots , n_k \rangle \in \text{Seq} \Rightarrow \text{lh}(\alpha) &= k.
\end{align*}

\item \textit{Конкатенация}~---~соединение двух последовательностей:
$$
\alpha = \langle m_1, \dots , m_p \rangle, \beta = \langle n_1, \dots , n_q \rangle \in \text{Seq} \Rightarrow a * b = \langle m_1, \dots , m_p, n_1, \dots , n_q \rangle.
$$

\item $\alpha$~---~начальный отрезок последовательности $\beta$:
$$
\alpha = \langle m_1, \dots , m_p \rangle, \beta = \langle n_1, \dots , n_q \rangle \in \text{Seq} \land p \leq q \land \forall i \leq p: m_i = n_i \Rightarrow \alpha \leq \beta.
$$
\end{itemize}

Пусть $D = (D, \sqsubseteq)$~---~частично упорядоченное множество с рефлексивным отношением $\sqsubseteq$.
Тогда подмножество $X \subseteq D$ называется \textit{направленным}, если
$$
X \neq \varnothing \land \forall x, y \in X: \exists z \in X: x \sqsubseteq z \land y \sqsubseteq z.
$$

При этом $D$ называется \textit{полным}, если для любого направленного подмножества $X \subseteq D$ существует \textit{супремум} $\sqcup X \in D$ и имеется \textit{дно} $\bot$:
$$
\exists \bot \in D: \forall x \in D: \bot \sqsubseteq x.
$$

\textit{Топология Скотта} на полном частично упорядоченном множестве $(D, \sqsubseteq)$ определяется следующим образом: множество $O \subseteq D$ считается \textit{открытым}, если выполняются два условия.

\begin{enumerate}
\item  $x \in O \land x \sqsubseteq y \Rightarrow y \in O$.
\item $X \subseteq D \land \sqcup X \in O \Rightarrow X \cap O \neq \varnothing$.
\end{enumerate}

\textit{Частичное отображение} $\phi: X \leadsto Y$~---~это отображение $\phi$, такое, что область определения $\text{Dom}(\phi) \subseteq X$.
Для $x \in X$ запись $\phi(x)\downarrow$ означает, что $\phi(x)$ определено, то есть $x \in \text{Dom}(\phi)$; $\phi(x)\uparrow$ означает, что $\phi(x)$ не определено, то есть $x \not\in \text{Dom}(\phi)$.

Если $\Sigma$~---~некоторое множество символов, то \textit{частично $\Sigma$-помеченное дерево}~---~это частичное отображение $\phi: \text{Seq} \leadsto \Sigma \times \mathbb N$, такое, что выполняются два условия.

\begin{enumerate}
\item $\phi(\sigma)\downarrow \land \tau \leq \sigma \Rightarrow \phi(\tau)\downarrow$.
\item $\phi(\sigma) = \langle a, n \rangle \Rightarrow \forall k \geq n: \phi(\sigma * \langle k \rangle)\uparrow$.
\end{enumerate}

\textit{Обнаженное дерево, лежащее в основе} частично $\Sigma$-помеченного дерева $\phi$,~—~это
$$
T_\phi = \{\langle \phantom M \rangle\} \cup \{\sigma\ |\ \sigma = \sigma' * \langle k \rangle \land \phi(\sigma') = \langle a, n \rangle \land k < n\}.
$$

Если $\sigma \in T_\phi$ и $\phi(\sigma) = \langle a, n \rangle$, то $a$ называется \textit{меткой} в узле $\sigma$.
Если же для $\sigma \in T_\phi$ $\phi(\sigma)\uparrow$, то говорят, что узел $\sigma$ \textit{непомеченный}.
Частично помеченные деревья будем обозначать заглавными буквами и будем писать $\sigma \in A$ вместо $\sigma \in T_A$ и $A(\alpha) = \bot$, когда $A(\alpha)\uparrow$, но все же $\alpha \in A$.

Если $\Sigma = \{\lambda x_1 \dots \lambda x_n.x\ |\ n \geq 0, x_1, \dots , x_n, x \in \Lambda\}$, то частично $\Sigma$-помеченное дерево называется \textit{деревом бемовского типа}.
Множество всех таких деревьев обозначим $B$.
\textit{Поддерево дерева $A$, исходящее из узла $\alpha$}~---~это $A_\alpha = \mathbold\lambda \beta.A(\alpha * \beta)$.
Очевидно, что $\forall A \in B: \forall \alpha: A_\alpha \in B$.

Комбинатор $M$ \textit{разрешим}, если
$$
\exists n: \exists N_1, \dots , N_n \in \Lambda^0: M\ N_1 \dots N_n = I.
$$
Например, комбинатор неподвижной точки разрешим, так как
$$
Y\ (K\ I) = K\ I\ (Y\ (K\ I)) = I.
$$
С другой стороны, $\Omega$ неразрешим.
Произвольное $\lambda$-выражение \textit{разрешимо}, если разрешим комбинатор $\lambda x_1 \dots \lambda x_n.M$, где $\{x_1, \dots , x_n\} = \text{FV}(M)$.

$\lambda$-выражение $M$ является \textit{головной нормальной формой}, если оно имеет вид
$$
M \equiv \lambda x_1 \dots \lambda x_n.x\ M_1 \dots M_m, \quad m, n \geq 0.
$$
Говорят, что $M$ \textit{имеет} головную нормальную форму $N$, если $M = N$.
\textit{Главной} называется та головная нормальная форма выражения, которая первой достигается его левой редукцией.

Уодсворт ввел класс $\lambda$-выражений, не имеющих головной нормальной формы, и привел доводы в пользу того, что элементы этого класса должны рассматриваться как бессмысленные выражения в $\lambda$-исчислении.
Ему принадлежит следующий важный результат: $\lambda$-выражение разрешимо тогда и только тогда, когда оно имеет головную нормальную форму.
Таким образом, из неразрешимости $M$ следует, что для любых выражений $N_1, \dots , N_n$ выражение $M\ N_1 \dots N_n$ не имеет нормальной формы.

\textit{Дерево Бема} для терма $M$, обозначаемое через $\text{BT}(M)$,~---~это дерево бемовского типа, определяемое следующим образом.

\begin{enumerate}
\item Если $M$ неразрешим, то $\forall \sigma: \text{BT}(M)(\sigma)\uparrow$.
\item Если $M$ разрешим и имеет главную головную нормальную форму
$$
\lambda x_1 \dots \lambda x_n.x\ M_0 \dots M_{m - 1},
$$
то дерево Бема определяется рекурсивно:
\begin{align*}
\text{BT}(M)(\langle \phantom M \rangle) &= \langle \lambda x_1 \dots \lambda x_n.x, m \rangle; \\
k < m \Rightarrow \text{BT}(M)(\langle k \rangle * \sigma) &= \text{BT}(M_k)(\sigma); \\
k \geq m \Rightarrow \text{BT}(M)(\langle k \rangle * \sigma)&\uparrow.
\end{align*}
\end{enumerate}

Рассмотрим полное частично упорядоченное множество $B = (B, \subseteq)$ с топологией Скотта.
\textit{Топология деревьев} на множестве $\Lambda$~---~это наименьшая топология, в которой непрерывно отображение $\text{BT}: \Lambda \rightarrow B$.
Иными словами, открытые подмножества $\Lambda$ имеют вид $\text{BT}^{-1}(O)$, где $O$ открыто в топологии Скотта на $B$.

Используя топологию деревьев, можно выразить обычные понятия, относящиеся к $\lambda$-исчислению, в топологических терминах.
Например, нормальные формы оказываются изолированными точками, а неразрешимые выражения — \textit{точками компактификации}, то есть такими точками, единственной окрестностью которых является само топологическое пространство.

Доказано, что аппликация и абстракция непрерывны в топологии деревьев на $\Lambda$, причем для аппликации это нетривиальный результат, имеющий интересные следствия.
Например, множество $\text{Sol} \subseteq \Lambda$ разрешимых термов открыто.
Действительно, в любом полном частично упорядоченном множестве множество $\{x\ |\ x \neq \bot\}$ открыто по Скотту.
Следовательно, множество $\text{Sol} = \text{BT}^{-1}\{A\ |\ A \neq \bot\}$ открыто в $\Lambda$.

\end{document}